\documentclass[12pt]{article}
\usepackage{amsmath}
\usepackage{amsfonts}

\usepackage{mltex}

\newtheorem{lemma}{Lemma}[section]
\newtheorem{theorem}[lemma]{Theorem}
\newtheorem{proposition}[lemma]{Proposition}
\newtheorem{corollary}[lemma]{Corollary}
\newtheorem{remark}[lemma]{Remark}

\newtheorem{definition}[lemma]{Definition}

\def\sq{\hbox {\rlap{$\sqcap$}$\sqcup$}}
\def\sq{\hbox {\rlap{$\sqcap$}$\sqcup$}}
\def\R{\mathbb R}
\def\C{\mathbb C}
\def\Z{\mathbb Z}
\def\1{{\rm 1\mskip-4.5mu l} }
\def\lsim{\raise0.3ex\hbox{$<$\kern-0.75em\raise-1.1ex\hbox{$\sim$}}}
\def\gsim{\raise0.3ex\hbox{$>$\kern-0.75em\raise-1.1ex\hbox{$\sim$}}}
\def\noi{\noindent}
\def\beq{\begin{equation}}   \def\eeq{\end{equation}}
\def\bea{\begin{eqnarray}}  \def\eea{\end{eqnarray}}

\def\noi{\noindent}
\def\di{\displaystyle}
\newcommand{\QED}{\mbox{}\hfill
\raisebox{-2pt}{\rule{5.6pt}{8pt}\rule{4pt}{0pt}}
            \medskip\par}

\renewcommand{\theequation}{\thesection.\arabic{equation}}
\newcounter{hran} \renewcommand{\thehran}{\thesection.\arabic{hran}}

\def\bmini{\setcounter{hran}{\value{equation}}
    \refstepcounter{hran}\setcounter{equation}{0}
    \renewcommand{\theequation}{\thehran\alph{equation}}\begin{eqnarray}}

\def\bminiG#1{\setcounter{hran}{\value{equation}}
\refstepcounter{hran}\setcounter{equation}{-1}
\renewcommand{\theequation}{\thehran\alph{equation}}
\refstepcounter{equation}\label{#1}\begin{eqnarray}}

\def\emini{\end{eqnarray}\relax\setcounter{equation}{\value{hran}}\renewcommand{
\theequation}
{\thesection.\arabic{equation}}}

\begin{document}

\title {Quadratic Quantum Hamiltonians revisited}
\author{\it {\bf Monique Combescure} \\
\it IPNL, B\^atiment Paul Dirac \\
\it 4 rue Enrico Fermi,Universit\'e Lyon-1 \\
\it  F.69622 VILLEURBANNE Cedex, France\\
\tt monique.combescure@ipnl.in2p3.fr\\
\\
\it {\bf Didier Robert} \\
\it D\'epartement de Math\'ematiques\\
\it Laboratoire Jean Leray, CNRS-UMR 6629\\
\it Universit\'e de Nantes, 2 rue de la Houssini\`ere, \\
F-44322 NANTES Cedex 03, France\\
\tt Didier.Robert@math.univ-nantes.fr}
\vskip 1 truecm
\date{}
\maketitle

\begin{abstract}
Time dependent quadratic Hamiltonians  are well known as well in classical mechanics and in quantum mechanics.
In particular  for them the correspondance between classical and quantum mechanics is exact.
 But explicit formulas are non trivial (like the Mehler formula). Moreover,  a good knowlege of quadratic Hamiltonians
  is very useful  in the study of more general  quantum Hamiltonians and associated  Schr\"odinger equations in the semiclassical regime.\\
 Our goal here is to  give our own presentation of this important subject. We put emphasis on computations  with Gaussian coherent states.
  Our main motivation to do that is application concerning revivals and Loschmidt echo.
\end{abstract}

\newpage

\section{Introduction}
This paper is a survey concerning exact useful formulas for 
 time dependent Schr\"odinger equations  with quadratic Hamiltonians in the phase space.
  One of our motivations  is to give a detailed proof for the computation of the Weyl symbol of the
   propagator. This formula was used recently by Melhig-Wilkinson \cite{mewi}  to suggest a simpler  proof of  the Gutzwiller trace formula \cite{coraro}.\\
     There exist  many  papers concerning  quantum quadratic Hamiltonians and exact
      formulas.  In 1926, Schr\"odinger \cite{sch}   has already remarked that quantification of the harmonic (or Planck)  oscillator is exact. \\
      The best known result  in this field is  certainly the Melher formula for the harmonic oscillator
     (see for example \cite{cfks}).
     
     Quadratic Hamiltonians are very  important in partial differential equations on one side because they give non trivial examples of wave propagation phenomena and in quantum mechanics and on  the otherside the propagation of coherent states by general classes
      of Hamiltonians, including $-\hbar^2\triangle + V$,  can be  approximate modulo
       $O(\hbar^\infty)$  by evolutions of quadratic time dependent Hamiltonians
        \cite{coro1, ro2, lasi}.

     In his works on pseudodifferential calculus,  A. Unterberger in \cite{un1, un2} has given  several
      explicit formulas connecting harmonic oscillators, Gaussian functions and the symplectic group.
      This subject was also studied in \cite{fo, how, ho2}. 
      More  recently de Gosson \cite{dg} has given  a different approch for a rigorous proof  of the Melhig-Wilkinson formula
 for metaplectic operators,  using  his previous  works on symplectic geometry and   the  metaplectic group. \\
 Here we shall  emphasis on time dependent quadratic Schr\"odinger equation
  and  Gaussian Coherent States. It is well known and clear 
   that this approach is in the heart of the subject and was 
   more or less present  in all papers on quantum quadratic Hamiltonians. 
   In this   survey, we want to give our own presentation of the subject
    and  cover most of results appearing  in  particular \cite{how, fed}.\\
    Our main  motivation to revisit this subject was to  prepare useful tools  for  applications to
     revivals and quantum Loschmidt echo \cite{coro2}.
   We shall see  that in our approach  computations are rather natural  direct and explicit.
  
\section{Weyl quantization. Facts and Notations}
Let us first recall some well known facts concerning Weyl quantization (for more details see
 \cite{ho1, ro1}). \\
 The Planck constant $\hbar>0$ is fixed (it is enough to assume $\hbar = 1$  in the homogeneous quadratic case).\\
 The Weyl quantization is a continuous linear map, denoted by $\hat\bullet$
  or  by $Op^w$, 
  defined on the temperate Schwartz space distribution ${\cal S}^\prime(\mathbb R^{2n})$
   into ${\cal L}_w({\cal S}(\mathbb R^{n}), {\cal S}^\prime(\mathbb R^{n}))$
    where ${\cal L}_w(E, F)$ denotes the linear space of continuous linear map from the
     linear topological $E$ into the linear topological $F$ with the weak topology.\\
    $\mathbb R^{2n}$ is a symplectic linear space with the canonical symplectic form
     $\sigma(X,Y)=JX\cdot Y$ where  $J$ is  defined by
          $$ 
     J = \left(\begin{array}{cc}
0&\1_{n}
\\
-\1_{n}&0
\end{array}\right) 
$$ 
and $\1_{n}$ denotes the identity $n \times n$ matrix.\\
Let us introduce   the symplectic group ${\rm Sp}(2n)$:  it is  the set of linear transformations of $\mathbb R^{2n}$
 preserving the 2-form $\sigma$.\\
For  $X\in\mathbb R^{2n}$, we denote  $X=(q,p)\in\mathbb R^n\times\mathbb R^n$.\\
The Weyl quantization is uniquely determined by the following conditions:\\

(W0) $A\mapsto \hat A$ is continuous.\\     

(W1) $(\hat q\psi)(q) = q\psi(q),\;\; (\hat p\psi)(q) = D_q\psi(q),\;
 (D_q=\frac{\nabla_q}{i}$)\\
for every $\psi\in{\cal S}(\mathbb R^n)$.\\

(W2)$\ \exp[i(\widehat{\alpha\cdot q+\beta\cdot p)}] = 
\exp[i(\alpha\cdot\hat q+\beta\cdot\hat p)], \qquad 
\forall \ \alpha, \beta \in \mathbb R^n$
\\

Let us remark that $\alpha\cdot\hat q+\beta\cdot\hat p$ is self-adjoint  on 
$L^2(\mathbb R^n)$
 so, $\exp[i(\alpha\cdot\hat q+\beta\cdot\hat p)]$ is unitary. In particular,
 if $z=(x,\xi)$ then 
 $$
 \hat T(z)=: \exp[\frac{i}{\hbar}(\xi\cdot\hat q-x\cdot\hat p)
 $$
 is the quantized translation  by $z$ in the phase space (Weyl operators).\\
 From (W1), (W2), using continuity and $\hbar$-Fourier transform, defined
  by
  $$
  \tilde{A}(Y) = \int_{\mathbb R^{2n}}{\rm e}^{-\frac{i}{\hbar}X\cdot Y}A(X)dX
  $$
  we have
\beq\label{Weyl1}
  \hat A = (2\pi\hbar)^{-2n}\int_{\mathbb R^{2n}}\tilde A(\alpha,\beta)
  \exp[\frac{i}{\hbar}(\alpha\cdot\hat q+\beta\cdot\hat p)]d\alpha d\beta
  \eeq 
  In general, equality (\ref{Weyl1}) is only defined in a weak sense i.e through the duality bracket between ${\cal S}^\prime$ and ${\cal S}$, 
   $<\hat A\varphi, \psi>$ for arbitrary $\varphi, \psi\in {\cal S}(\mathbb R^n)$.\\
   
   \begin{definition}
   $A$ is  the Weyl contravariant symbol of $\hat A$ if they are  related through the formula
   (\ref{Weyl1}).
      \end{definition}
      
      Using the explicit action of Weyl operators on ${\cal S}(\mathbb R^n)$ and Fourier
       analysis, we get the following formula (see \cite{ho1, ro1})
       \beq\label{Weyl2}
       (\hat A\varphi)(x) = 
       (2\pi\hbar)^{-n}\int\!\!\int_{\mathbb R^{2n}}\exp[\frac{i}{\hbar}(x-y)\cdot\xi]
       A\left(\frac{x+y}{2},\xi\right)\varphi(y)dyd\xi.
       \eeq
       In particular, if $K_A$ denotes the Schwartz kernel of $\hat A$, we have
       \beq\label{Weyl3}
       A(q,p) = \int_{\mathbb R^n}\exp[-\frac{i}{\hbar}u\cdot p]
       K_{A}\left(q+\frac{u}{2}, q-\frac{u}{2}\right)du
       \eeq
     These formulas are true in the distribution sense in general, and pointwise if
      $\hat A$ is smoothing enough ( surely if for example  $A\in{\cal S}(\mathbb R^n)$).\\
   Let us remark that they are consistent and that $A\mapsto \hat A$ is a bijection
    from ${\cal S}^\prime(\mathbb R^{2n})$
   into ${\cal L}_w({\cal S}(\mathbb R^{2n}), {\cal S}^\prime(\mathbb R^{2n}))$.
   In particular we have the  following inversion formula
   \begin{proposition}
   For every $\hat A\in {\cal L}_w({\cal S}(\mathbb R^{2n}), {\cal S}^\prime(\mathbb R^{2n}))$,
    there exists a unique  contravariant Weyl symbol $A\in {\cal S}^\prime(\mathbb R^{2n})$ 
     given by the following formula
     \beq\label{Weyl4}
     A(X) = 2^n{\rm Tr}[\hat A S_{ym}(X)]
     \eeq
     where $S_{ym}(X)$ is the unitary operator in $L^2(\mathbb R^n)$ defined by
     $$
      S_{ym}(X)\varphi(q)= 
      (\pi\hbar)^{-n}{\rm e}^{\frac{-2i\xi}{\hbar}(x-q)}\varphi(2x-q)
      $$
    for  $X=(x,\xi)$.
     \end{proposition}
     
     {\bf Sketch of proof}:\\
     
     We first prove the formula  for
      $\hat A\in {\cal L}_w({\cal S}^\prime(\mathbb R^{2n}), 
      {\cal S}(\mathbb R^{2n}))$.The general case follows by duality and density.\\
     We start with the following  formula, easy to prove if 
     $A, B\in {\cal S}(\mathbb R^{2n})$, 
     \beq\label{trace}
     {\rm Tr}[\hat A\hat B] =(2\pi\hbar)^{-n} \int_{\mathbb R^{2n}}A(X)B(X)dX
     \eeq
     Assume now that $\hat B$ is a bounded operator in $L^2(\mathbb R^n)$.
       The Weyl symbol $B$
      of $\hat B$ satisfies:
   \beq\label{dualtrace}
      {\rm Tr}[\hat A\hat B] =(2\pi\hbar)^{-n} <A, B>\footnote{the  bracket  $<,>$ denotes
       the  usual bilinear form (integral or distribution pairing).  We shall denote $\langle\bullet\vert\bullet\rangle$  the Hermitean  sesquilinear form on  Hilbert spaces, linear in the second argument.}
\eeq
   we have to check that the Weyl symbol of  $ S_{ym}(X)$
    is $(\pi\hbar)^n\delta_X$ where $\delta_X$ is the Dirac mass in $X$.\\
    To prove that let us consider any $\varphi\in{\cal S}(\mathbb R^n)$ and denote
     by $W_\varphi$ the Weyl symbol of the projector $\psi\mapsto <\psi,\varphi>
     \varphi$. 
      $W_\varphi$  is called the Wigner function of $\varphi$. A direct computation  using
       (\ref{Weyl3}) gives
      $$
      W_\varphi(q,p) = \int_{\mathbb R^n}\exp[-\frac{i}{\hbar}u\cdot p]
      \varphi(q+\frac{u}{2})
      \overline{\varphi( q-\frac{u}{2})}du
    $$
     We get the  proposition  applying formula (\ref{dualtrace}) with 
  $\hat B=S_{ym}(X)$ and $\hat A=\pi_\varphi$.\sq
  
  We shall see later that it may be convenient to introduce the covariant Weyl  symbol 
   for $\hat A$ which has a nice connection with Weyl translations.
   
   \begin{proposition}
   For every 
 $\hat A\in {\cal L}_w({\cal S}(\mathbb R^{2n}), {\cal S}^\prime(\mathbb R^{2n}))$
 there exists a unique temperate distribution $A^\#$ on $\mathbb R^{2n}$,
  named covariant Weyl symbol of $\hat A$, 
    such that
    \beq
    \hat A = (2\pi\hbar)^{-n}\int_{\mathbb R^{2n}}A^\#(X)\hat T(X)dX
    \eeq
    Moreover we have the inverse formula
    \beq
    A^\#(X) ={\rm Tr}[\hat A\hat T(-X)]
        \eeq
       ( As  above, if $\hat A$ is not trace-class, this formula  has to be interpreted in 
         a weak distribution  sense).\\
         The covariant and contravariant Weyl symbols  are related with the following formula
         \beq
         A^\#(X) = (2\pi\hbar)^{-n}\tilde{A}(JX). 
         \eeq
         $\tilde{A}\circ J$ is named the symplectic Fourier transform of $A$
   \end{proposition}
   
   {\bf Proof}\\   These properties are not difficult to prove
    following  for example \cite{lit}.\sq\\
    
    We define the (usual) Gaussian coherent states $\varphi_{z}$ as follows:
    \beq
    \varphi_{z}= \hat T(z)\varphi_{0}, \qquad \forall\ z\in\mathbb R^{2n}
    \eeq
    where
    \beq
    \varphi_{0}(x):= (\pi \hbar)^{-n/4}e^{-x^2/2\hbar}
    \eeq
    
    We get the following useful formula for the mean value of observables:
    
    \begin{corollary} 
    With the above notations, for every $\varphi, \psi\in{\cal S}(\mathbb R^n)$,
     we have
    \beq\label{mean}
    \langle\varphi\vert\hat A\psi\rangle = (2\pi\hbar)^{-n}\int_{\mathbb R^{2n}}
    A^\#(X)\langle\varphi\vert\hat T(X)\psi\rangle dX
    \eeq  
    In particular for Gaussian Coherent States, we have
    \beq\label{meanGauss}
    \langle\varphi_z\vert\hat A\varphi_0\rangle =
    (2\pi\hbar)^{-n}\int_{\mathbb R^{2n}}A^\#(X)
    \exp\left(\frac{-\vert X-z\vert^2}{4\hbar}-\frac{i}{2\hbar}\sigma(X,z)\right)dX
    \eeq
    \end{corollary}
    
    {\bf Proof}:\\
The first   formula is a direct consequence of definition for covariant symbols.\\
The second formula is a consequence of the first and the following easy to prove equalities
\bea
\hat T(z)\hat T(z^\prime) &=& \exp\left(\frac{i\sigma}{2\hbar}(z,z^\prime)\right)\hat T(z+z^\prime)\\
\langle\varphi_z\vert\varphi_0\rangle &=& {\rm e}^{-\frac{\vert z\vert^2}{4\hbar}}
\eea
  \sq

For later use let us recall the following
\begin{definition}
Let be $\hat A, \hat B\in {\cal L}_w({\cal S}(\mathbb R^{n}), {\cal S}^\prime(\mathbb R^{n}))$
 such that the operator composition  $\hat A\hat B$ is well defined.
  Then the Moyal product of $A$ and $B$ is
  the unique $A\#B\in{\cal S}^\prime(\mathbb R^{2n})$ such that
   \beq 
   \hat A\hat B = \widehat{A\# B}
   \eeq
\end{definition}
For details concerning  computations  rules and properties of Moyal products
 see \cite{ho1, ro1}.

       \section{Time evolution of Quadratic Hamiltonians}
In this section we consider a  quadratic  time-dependent Hamiltonian,  \\
 $H_t(z) =\sum_{1\leq j, k \leq 2n}c_{j,k}(t)z_jz_k$, with real and continuous coefficients $c_{j,k}(t)$,
  defined on the whole real line for simplicity. It is convenient to consider the symplectic
   splitting  $z = (q, p)\in \mathbb R^{n}\times \mathbb R^n$
   and  to write down  $H_t(z)$ as
   $$
   H_t(q,p) = \frac{1}{2}\left(G_tq\cdot q + 2L_tq\cdot p + K_tp\cdot p\right)
   $$ 
   where  $K_t, L_t, G_t$ are real $n\times n$  matrices, $K_t$ and $G_t$ being symmetric.\\ The classical motion in the phase space
    is given by the linear equation
    \beq\label{classicevol}
    \left(\begin{array}{c}\dot q\\ \dot p\end{array}\right) 
     = J.\left(\begin{array}{cc}G_t & L_t^T\\
     L_t & K_t \end{array}\right)\left(\begin{array}{c}q \\p\end{array}\right),
     \eeq
     where $L^T$ is the transposed matrix of $L$. 
     This equation defines a flow, $F_{t}$ (linear symplectic transformations)
      such that  $F_{0} = \1$.
      On the quantum side, $\widehat{H_t}$ is a family of self-adjoint operators on the Hilbert space
       ${\cal H} = L^2(\mathbb R^n)$ (this will be proved later). The quantum evolution follows the
        Schr\"odinger equation, starting with an initial state $\varphi\in{\cal H}$.
        \beq\label{quantevol}
        i\hbar\frac{\partial \psi_t}{\partial t} = \widehat{H_t}\psi_t,\;\;  \psi_{t_0} = \varphi
        \eeq
        Suppose that we have  proved existence and  uniqueness  for solution of 
        (\ref{quantevol}), we write
         $\psi_t = \widehat{U_{t}}\varphi$. The correspondence between the classical evolution
          and quantum evolution is exact. For every $A\in{\cal S}(\mathbb R^{2n})$, we have 
          \begin{proposition}\label{Egexact}
          \beq
          \widehat{ U_{t}}.\widehat{A}.\widehat{U_{t}^{-1}} = \widehat{A.F_{t}}
           \eeq
           \end{proposition}
           
        \noi {\bf Sketch of proof}\\
         For any quadratic Weyl symbol $B$ we have the exact formula
         \beq\label{Poisson}
         \frac{i}{\hbar}[\widehat B, \widehat A] = \widehat{\{B, A\}}
         \eeq
          where  ${\{B, A\}}=\nabla B\cdot J\nabla A$  denote the Poisson bracket,
          and $[\hat B, \hat A]=\hat B.\hat A - \hat A.\hat B$ is the Moyal bracket.
          So we can prove Proposition  (\ref{Egexact}) by taking derivative in time
           and using  (\ref{Poisson}) (see \cite{ro1} for  more details).
          \QED
          Now we want to compute explicitly the quantum propagator  $U_{t_0,t}$
           in terms of classical evolution of $H_t$. \\
         One  approach  is to   compute  the time evolution of
           Gaussian coherent sta(es,  $\widehat U_{t_0,t}\varphi_z$,   or in other word to solve the
            Schr\"odinger equation (\ref{quantevol}) with
            $\varphi = \varphi_z$, the Gaussian coherent state in $z\in \mathbb R^n$.   
            Let us recall that $\varphi_z=\hat T(z)\varphi_0$ and
             $\varphi_0(x) = (\pi\hbar)^{-n/4}\exp\left(\frac{-\vert x\vert^2}{2\hbar}
             \right)$.
             
      \section{Time evolution of Coherent States}
      The coherent states  system $\{\varphi_z\}_{z\in \R^{2n}}$  introduced before is a very convenient tool to analyze properties of operators in $L^2(\R^n)$ and their Schwartz distribution kernel. To understand that let us underline  the following consequence of the Plancherel Formula for the Fourier transform.
       In all this section we assume $\hbar = 1$.  For every $u\in L^2(\R^n)$ we have
\beq\label{fbp}
      \int_{\R^n}\vert u(x)\vert^2dx = (2\pi)^{-n}\int_{\R^{2n}}\vert \langle u, \varphi_z\rangle\vert^2dz
  \eeq
 Let  $\hat R$  be some continuous linear opertor from ${\cal S}(\R^n)$ into     ${\cal S}^\prime(\R^n)$  and $K_R$ its 
  Schwartz distribution kernel. By an easy computation, we get the following representation formula
  \beq\label{sck}
  K_R(x, y) = (2\pi)^{-n}\int_{\R^{2n}}(\hat R\varphi_z)(x)\overline{\varphi_z(y)}dz.
  \eeq
  In other words  we have the following  continuous resolution of the   Schwartz distribution kernel of the identity
  $$
  \delta(x-y) = (2\pi)^{-n}\int_{\R^{2n}}\varphi_z(x)\overline{\varphi_z(y)}dz.
  $$
This formula explains why the Gaussian coherent system  may be an efficient tool for analysis of operators on 
 the Euclidean space $\R^n$.\\
    Let us  consider first  the harmonic oscillator
      \beq
\hat H = - \frac{1}{2}\frac{d^2}{dx^2} + \frac{1}{2}x^2
\eeq
It is well known that for  $t \not= k\pi, \ k \in \Z$ the quantum propagator  ${\rm e}^{-it \hat H}$  has an  explicit  Schwartz
 kernel $K(t; x, y)$  (Mehler formula).
\\
It is easier to  compute with the coherent states $\varphi_{z} $.
 $\varphi_0$ is an  eigenstate  of $\hat H$, so we have
\beq
{\rm e}^{-it\hat H}\varphi_{0} =  {\rm e}^{-it/2}\varphi_{0}
\eeq
Let us compute ${\rm e}^{-it\hat H}\varphi_{z}$, $ \forall z \in \R^2$, with the following ansatz
\beq\label{ansatz1}
{\rm e}^{-it\hat H}\varphi_{z}= {\rm e}^{i\delta_{t}(z)}\hat T(z_{t})e^{-it/2}\varphi_{0}
\eeq
where  $z_{t}= (q_{t}, p_{t})$  is the generic point on the classical trajectory (a circle here), coming from $z$ at time $t= 0$. 
Let be  $\psi_{t,z}$ the state equal to the r.h.s in (\ref{ansatz1}), and let us compute $\delta_{t}(z)$  such that  $\psi_{t, z}$ satisfies the equation $i\frac{d}{dt}\varphi = \hat H \varphi \quad \varphi \vert_{t=0}= \psi_{0, z}$. 
We have 
$$\hat T(z_{t})u(x)= e^{i(p_{t}x - q_{t}p_{t}/2)}u(x-q_{t})$$
and  
\beq\label{ansatz2}
\psi_{t,z}(x)= {\rm e}^{i(\delta_{t}(z) -t/2 + p_{t}x - q_{t}p_{t}/2)}\varphi_{0}(x-q_{t})
\eeq
So, after some computations left to the reader, using properties of the classical trajectories
$$
\dot q_{t} = p_{t} \quad \dot p_{t} = - q_{t}, \quad p_{t}^2 + q_{t}^2 = p^2 + q^2,
$$
the equation 
   \beq\label{ansatz3}
   i \frac{d}{dt}\psi_{t,z}(x)= \frac{1}{2}(D_{x}^2 + x^2)\psi_{t, z}(x)
   \eeq
is satisfied if and only if 
\beq
\delta_{t}(z)= \frac{1}{2}(p_{t}q_{t} - pq)
\eeq
\sq

          Let us  now introduce the following  general notations for later use.\\
          $F_{t}$  is the classical flow with initial time $t_0=0$ and final time $t$. It is 
          represented as a     $2n\times 2n$ matrix  
  which can be written as four $n\times n$ blocks~:
 \beq 
F_{t}= \left(\begin{array}{cc}A_{t} & B_{t }\\
C_{t}&  D_{t} \end{array}\right).
\eeq
Let us introduce the following  squeezed states.
    $\varphi^{\Gamma}$  is  defined as follows.
        \beq
    \varphi^{\Gamma}(x) =  a_\Gamma
    \exp\left(\frac{i}{2\hbar}\Gamma x\cdot x\right)
    \eeq
   where     $\Gamma\in\Sigma_n$, $\Sigma_n$  is  the  Siegel space of complex, symmetric
    matrices $\Gamma$ such that $\Im(\Gamma)$ is positive and non degenerate
     and $a_\Gamma\in\mathbb C$ is such that the $L^2$-norm of $  \varphi^{\Gamma}$ is one.\\
     We also denote $ \varphi^{\Gamma}_z = \widehat{T}(z) \varphi^{\Gamma}$.\\
     For $\Gamma = i\1$, we denote $\varphi = \varphi^{i\1}$.
       \begin{theorem}\label{quadpropag1}
 We have the following formulae, for every  $x\in \mathbb R^n$ and $z\in \mathbb R^{2n}$, 
        \bea\label{propagauss}
            \widehat U_{t}\varphi^\Gamma (x) &=& \varphi^{\Gamma_t}(x) \\
             \widehat U_t\varphi^\Gamma_z (x) &=& \widehat{T}(F_{t}z)\varphi^{\Gamma_t}(x) 
               \eea
                    where      $\Gamma_{t} = (C_{t} + iD_{t}\Gamma)(A_{t} + iB_{t}\Gamma)^{-1}$
              and    $a_{\Gamma_t} = a_\Gamma\left({\rm det}(A_{t}+iB_{t}\Gamma)\right)^{-1/2}$.
           \end{theorem}
          \noi{\bf Beginning of the  proof}\\
                 The first formula can be proven by the ansatz
                                  $$
 \widehat U_{t}\varphi_0 (x) = a(t)\exp\left(\frac{i}{2\hbar}
  \Gamma_tx\cdot x\right)
              $$
 where $\Gamma_t\in\Sigma_n$  and 
       $a(t)$ is a complex values time dependent function. We  get  first a Riccati equation to compute $\Gamma_t$ and a  linear
        equation to compute $a(t)$. \\
    The second formula is easy to  prove  from the first,  using the Weyl translation operators
     and the following  known property
     $$
    \widehat U_t\widehat{T}(z)\widehat U_t^* = \widehat{T}(F_tz).
     $$
    Let us now give the details of the proof for $z=0$. \\
    We begin by computing  the action of a quadratic Hamiltonian on a Gaussian ($\hbar$ = 1).
    
    \begin{lemma}
$$
Lx\cdot D_{x}{\rm e}^{\frac{i}{2}\Gamma x\cdot x} = ( L^Tx\cdot \Gamma x - \frac{i}{2}
{\rm Tr} L){\rm e}^{\frac{i}{2}\Gamma x\cdot x}
$$
\end{lemma}
 {\bf Proof}\\
This  is a straightforward computation, using
 $$
 Lx\cdot D_{x}  = \frac{1}{i}\sum_{1\leq j, k \leq n}L_{jk}\frac{x_{j}D_{k} + D_{k}x_{j}}{2}
 $$
and, for $\omega\in\R^n$, 
 $$(\omega . D_x){\rm e}^{\frac{i}{2}\Gamma x\cdot x} = 
  (\Gamma x\cdot \omega){\rm e}^{\frac{i}{2}\Gamma x\cdot x}
  $$
 \sq
 
 \begin{lemma}
 $$
 (G D_{x}\cdot D_{x}){\rm e}^{\frac{i}{2}\Gamma x\cdot x} =  
\left(G \Gamma x\cdot \Gamma x  - i{\rm Tr}(G \Gamma)\right){\rm e}^{\frac{i}{2}
 \Gamma x\cdot x}
 $$
 \end{lemma}
 
\noi
 {\bf Proof}\\
 As above, we get 
\beq
\widehat H{\rm e}^{\frac{i}{2}\Gamma x\cdot x} = 
  \left( \frac{1}{2}Kx\cdot x  +   x\cdot L \Gamma x  + \frac{1}{2}G \Gamma x\cdot \Gamma x  -\frac{i}{2 }{\rm Tr}(L + G \Gamma) \right){\rm e}^{\frac{i}{2} \Gamma x\cdot x}
\eeq

We are now ready to solve the equation 
\beq
i\frac{\partial}{\partial t}\psi = \hat H \psi
\eeq
with  $$\psi \vert_{t=0}(x) = g(x) := (2\pi)^{-n/2} e^{-x^2/2}.$$\\
We try the ansatz
\beq
\psi(t, x) = a(t)e^{\frac{i}{2} \Gamma_{t} x\cdot x}
\eeq
which gives the equations
\bea
\dot \Gamma_{t} &=& - K - 2\Gamma_{t}^TL - \Gamma_{t} G \Gamma_{t}\\
\dot f(t) &=& - \frac{1}{2}\left( Tr (L + G \Gamma_{t})\right)f(t)
\eea
with the initial conditions 
$$\Gamma_{0} = i \1, \quad a(0)= (2\pi)^{-n/2}$$

\noi{\bf Remark}: $ \Gamma^T L$ et $L \Gamma$ determine the same quadratic forms. 
So the first equation is a Ricatti equation  and can be written as
\beq\label{Ricatti}
\dot \Gamma_{t} = - K - \Gamma_{t} L^T - L \Gamma_{t} - \Gamma_{t }G \Gamma_{t},
\eeq
where $L^T$ denotes the transposed matrix for $L$.
We shall now see that equation (\ref{Ricatti}) can be solved using Hamilton equation
\bea
\dot F_{t} &=& J \left(
\begin{array}{cc}
K& L\\
 L^T& G
\end{array}
\right) F_{t}\\
F_{0} &=& \1
\eea
We know that $$F_{t} = \left(
\begin{array}{cc}
A_{t}& B_{t}\\
C_{t}& D_{t}
\end{array}
\right)$$
is a  symplectic matrix  $ \forall t$. So we have $\det( A_{t}+ i B_{t})\not= 0 \quad \forall t$ (see below). Let us denote
\beq
M_{t} = A_{t} + i B_{t}, \quad N_{t}= C_{t} + i D_{t}
\eeq
We shall prove that  $\Gamma_{t} = N_{t}M_{t}^{-1}$. 
By an easy computation, we get
\bea
\dot M_{t} &=&  L^T M_{t} + G N_{t}\nonumber\\
\dot N_{t}&=& -K M_{t} - LN_{t}
\eea
 
 Now, compute
 \bea
\frac{d}{dt}(N_{t}M_{t}^{-1}) &=& \dot N M^{-1} - N M^{-1 }\dot M M^{-1}\nonumber\\
&=& -K -LNM^{-1}-NM^{-1}(L^T M + GN)M^{-1}\nonumber\\
&=& -K -LNM^{-1 } -NM^{-1}  L^T - NM^{-1}GNM^{-1}
\eea
which is exactly equation (\ref{Ricatti}).\\
Now we compute $a(t)$, using the following equality, 
$$
{\rm Tr}\left( L^T + G(C+iD)(A+iB)^{-1}\right) = {\rm Tr}(\dot M)M^{-1}= {\rm Tr}\left(L + G \Gamma_{t}\right)$$
using ${\rm Tr}L = {\rm Tr}L^T$. Let us recall the Liouville formula
\beq
\frac{d}{dt}\log(\det M_{t})= {\rm Tr}(\dot M_{t}M_{t }^{-1})
\eeq
which give directly
\beq
a(t)= (2\pi)^{-n/2}\left( \det(A_{t}+i B_{t})\right)^{-1/2}
\eeq

To complete the proof,  we need to prove the following
\begin{lemma}
Let  $S$ be  a symplectic matrix.
$$S = \left(
\begin{array}{cc}
A&B\\
C&D
\end{array}
\right)$$
Then  $\det (A+iB)\not= 0$ and $\Im (C+iD)(A+iB)^{-1}$ is positive definite.
\end{lemma}

We shall prove a more general result concerning the  Siegel space $\Sigma_{n}$.

\begin{lemma} If 
$$S = \left(
\begin{array}{cc}
A&B\\
C& D
\end{array}
\right)$$ is a symplectic matrix and $Z\in\Sigma_n$  then A+BZ et C+DZ
are non singular   and 
$(C+DZ)(A+BZ)^{-1}\in \Sigma_{n}$
\end{lemma}
{\bf Proof}\\
Let us denote $E := A+BZ, \quad F:= C+DZ$.
 $F$ is symplectic, so we have  $ F^TJF= J$. Using
  $$\left(
\begin{array}{c}
E\\
F 
\end{array}
\right) = S 
\left(
\begin{array}{c}
I\\
Z
\end{array}
\right)$$
we get 
\beq
(E^T, F^T)J \left(
\begin{array}{c}
E\\
F
\end{array}
\right) = (I, Z)J \left(
\begin{array}{c}
I\\
Z
\end{array}
\right) = 0
\eeq
which gives
$$E^TF = F^T E$$
In the same way, we have
\beq
\frac{1}{2i}( E^T, F^T)J \left(
\begin{array}{c}
\bar E\\
\bar F\end{array}
\right) = \frac{1}{2i}(I, Z) F^T J F \left(
\begin{array}{c}
I\\
\bar Z\end{array}
\right)
\eeq
$$= \frac{1}{2i}(I, Z)J \left(
\begin{array}{c}
I\\
\bar Z
\end{array}
\right) = \frac{1}{2i}(\bar Z - Z)= -\Im Z$$
We get the following equation
\beq\label{inj}
F^T \bar E - E^T \bar F = 2i \Im Z
\eeq
Because $\Im Z$ is non degenerate, from (\ref{inj}), we get that $ E$ and  $F$ are injective.
 If  $x \in \C^n,\; Ex = 0$,  we have
$$\bar E \bar x = x^T  E^T= 0$$
hence $$ x^T \Im Z \bar x = 0$$ then $x =0$.\\
So, we can define, 
\beq
\alpha(S)Z = (C+DZ)(A+BZ)^{-1}
\eeq
Let us prove that  $\alpha(S) \in \Sigma_{n} $. We have:
$$
\alpha(M)Z = F E^{-1}\ \Rightarrow({\alpha(M)Z})^T = (E ^{-1})^T F^T = 
(E ^{-1 })^TE^TF E^{-1} =F E^{-1} = \alpha(M)Z.
$$
We have also:
$$ 
E^T \frac{F E ^{-1} - \bar F \bar E ^{-1}}{2i}\bar E = \frac { F^T \bar E -  E^T \bar F}{2i} = \Im Z
$$
and this proves that  $ \Im(\alpha(M)) $ is positive and non degenerate.\\
This finishes the proof of the Theorem for $ z=0$ . 
\sq

  \begin{remark}  For a different proof  of formula  (\ref{propagauss}),  using the usual approach of the metaplectic
   group, see the 
  book \cite{fo}.\\
  The family $\{\varphi_z\}_{z\in \mathbb R^{2n}}$ spans all of $L^2(\mathbb R^n)$
 (see for exemple \cite{ro2} for properties of the Fourier-Bargmann transform) so  formula (\ref{propagauss}) wholly determines  the unitary group $\widehat U_{t}$. In particular it results
   that $\widehat{U_{t}}$  is a unitary operator and that $\hat H_t$
   has a unique self-adjoint extension in $L^2(\mathbb R^n)$. This is left as exercises for the reader.
 \end{remark}   
                  
   \begin{remark}
   The map $S\mapsto\alpha(S)$ defines a representation of the symplectic group
    ${\rm Sp}(2n)$ in  the Siegel space $\Sigma_n$. It is easy to prove that 
         $\alpha(S_1S_2) = \alpha(S_1)\alpha(S_2)$. This representation is transitive.
      Many other properties of this representation are studied in \cite{ma}.
    \end{remark}
                            
           \section{The metaplectic group and  Weyl symbols computation}
A metaplectic transformation associated with a linear  symplectic tranformation $F\in{\rm Sp}(2n)$
 in $\mathbb R^{2n}$,  is a unitary operator
 $\widehat{R}(F)$ in $L^2(\mathbb R^n)$   satisfying one of the following equivalent conditions
            \bea
            \widehat{R}(F)^*\widehat{A}\widehat{R}(F)& =& \widehat{A\circ F},\;\;
             \forall A\in{\cal S}(\mathbb R^{2n})\\
             \widehat{R}(F)^*\widehat T(X)\widehat{R}(F)& = &\widehat T[F^{-1}(X)],\;\; \forall X\in\mathbb R^{2n}\\
             \widehat{R}(F)^*\widehat{A}\widehat{R}(F) &=&\widehat{A\circ F}, \nonumber\\
             {\rm  for}\; A(q,p)= q_j, 1\leq j\leq n  &{\rm and}&  
        A(q,p) = p_k, 1\leq k\leq n .        
            \eea
           We shall see below that for every $F\in{\rm Sp}(2n)$ there exists a metaplectic transformation $\widehat R(F)$.\\
            Let us remark that if $\widehat{R_1}(F)$ and $\widehat{R_2}(F)$ are two metaplectic operators associated to the
             same symplectic map $F$ then there exists $\lambda\in\mathbb C$, $\vert\lambda\vert=1$,
              such that  $\widehat{R_1}(F) = \lambda\widehat{R_2}(F)$.      
           It is also required  that  $F\mapsto \widehat{R}(F) $  defines  a projective representation
              of  the real symplectic group ${\rm Sp}(2n)$ with sign indetermination only.  More precisely,
               let us denote by ${\rm Mp}(n)$ the group of metaplectic transformations and $\pi_p$ the natural
                projection: ${\rm M_p}\rightarrow {\rm Sp}(2n)$ then the metaplectic representation  is a group homomorphism
                 $F\mapsto\hat{R}(F)$, 
                from ${\rm Sp}(2n)$ onto ${\rm Mp}(n)/\{\1,-\1\}$,  such that
                $\pi_p[\widehat{R}(F)]=F$,  $\forall F\in{\rm Sp}(2n)$
                (for  more details concerning the metaplectic representation
                 see \cite{le}).
                  We shall show here  that this can be  achieved straightforward  using Theorem 
               \ref{quadpropag1}.        \\
               
               For every $F\in {\rm Sp}(2n)$ we can find a  $C^1$- smooth curve  $F_t$, $t\in
                [0, 1]$, in      ${\rm Sp}(2n)$, such that
                $F_0 = \1$ and $F_1 = F$. 
                An explicit  way to do that is to use the polar decomposition of $F$, $F= V\vert F\vert$
                 where $V$ is a symplectic orthogonal matrix and $\vert F\vert = \sqrt{F^TF}$
                  is positive symplectic matrix. Each of these matrices have a logarithm, 
                   so $F= {\rm e}^K{\rm e}^L$ with $K, L$ Hamiltonian matrices,
                    and  we can choose  $F_t = {\rm e}^{tK}{\rm e}^{tL}$.
             Any way,    $F_t$ is clearly  the linear flow defined by the quadratic Hamiltonian
                 $H_t(z) = \frac{1}{2}S_tz\cdot z$ where $S_t = -J\dot{F_t}F_t^{-1}$. So using above results,
                  {\bf we define} $\widehat{R}(F)  = \widehat{U_1}$. From this definition
                   and Theorem \ref{quadpropag1} we can easily recover the usual properties of
                    the metaplectic representation.
                    
       \begin{proposition}
Let us consider two symplectic paths  $F_t$
                    and $F_t^\prime$ joining $\1\ (t=0)$  to $F\ (t=1)$. Then we have 
                    $\widehat{U_1} = \pm\widehat{ U_1^\prime}$ (with obvious notations). \\
                    Moreover, if $F^1, F^2\in{\rm Sp}(2n)$ then we have
                    $$
                    \widehat R(F^1)\hat R(F^2) = \pm\widehat R(F^1F^2).
                    $$
        \end{proposition} 
        
        \noi{\bf Proof}\\
        Using (\ref{propagauss}) we see that the  phase shift between the two paths comes
         from variation of argument between 0 and 1 
         of  the complex numbers $b(t) = {\rm det}(A_t+iB_t)$ and 
         $b^\prime(t) =  {\rm det}(A^\prime_t+iB^\prime_t)$.\\
         We have $\arg[b(t)]= \Im\left(\int_0^t\frac{\dot b(s)}{b(s)}
         ds\right)$ and
          it is well known (see  Lemma (\ref{index}) below and its proof) that
          $$
          \Im\left( \int_0^1\frac{\dot b(s)}{b(s)}ds\right)=  \Im
          \left(\int_0^1\frac{\dot b^\prime(s)}{b^\prime(s)}ds\right)+
           2\pi N
          $$
          with $N\in\mathbb Z$. So we get
          $$
          b(1)^{-1/2} = {\rm e}^{iN\pi}b^\prime(1)^{-1/2}
          $$
          The second part of the proposition is an easy consequence of Theorem \ref{quadpropag1}
           concerning propagation of  squeezed coherent states with little computations.
          \sq\\

 In a recent paper \cite{mewi} the authors use a nice explicit formula for the Weyl symbol
  of  metaplectic operators $\hat R(F)$. In what follows we detail a rigorous proof of this formula
   including  computation of the phase factor.           
In principle we could  use Theorem \ref{quadpropag1}  to compute the Weyl symbol of the
 propagator $U_{t_0,t}$.  But in this approach it seems difficult to compute phase factors 
 (Maslov- Conley-Zehnder  index).\\
  For technical reason, It is easier  for us to compute first the contravariant Weyl symbol,
$U_t$,  for the propagator $\widehat{U_t}$ 
   defined by $\widehat{H}_t$.  In any case, $U_t$ is  a Schwartz temperate distribution
    on the phase space $\mathbb R^{2n}$.\\
   We follow the approach used in Fedosov \cite{fed}.\\
   It is enough to assume  $\hbar = 1$.  In a first step we shall  solve the following problem
   \bea
   i\frac{\partial}{\partial t}\widehat{U^\varepsilon_t} &=& \widehat{H}_t\widehat{U^\varepsilon_t }\nonumber\\
   \widehat{U^\varepsilon_0} &=& \widehat{\1^\varepsilon}
   \eea
    where $\1^\varepsilon$ is a smoothing family of operators such that  
    $\displaystyle{\lim_{\varepsilon \rightarrow 0}\1^\varepsilon = \1}$. 
     It will be  convenient to take
     $\1^\varepsilon(X) = \exp(-\varepsilon\vert X\vert^2)$.\\
     
     Let us recall that \# denotes the Moyal product for Weyl symbols. So for the contravariant symbol $U_t(X)$ 
      of $U_t$ we have
     \beq
     i\frac{\partial}{\partial t}U_t(X) = (H_t\#U_t)(X)
     \eeq
     Because $H_t$ is a quadratic polynomial   we have
     \bea
     (H_t\#U_t)(X) = H_t(X)U_t(X) +\frac{1}{2i}\{H_t, U_t\}(X)  \nonumber\\
     -\frac{1}{8}(\partial_\xi\partial_y - \partial_x\partial_\eta)^2H_t(X)U_t(Y)\vert_{X=Y}
     \eea
     where $\partial_x = \frac{\partial}{\partial_x}$, $X=(x,\xi)$, $Y=(y,\eta)$.  
       \\
     It seems natural  to make the following  ansatz
     \bea\label{ans}
     U_t(X) &=& \alpha(t)E_t(X),\; {\rm where} \nonumber\\
     E_t(X) &=&  \exp\left(iM_tX\cdot X\right).
     \eea
    $\alpha(t)$ is a complex time dependent function, $M_t$ is a time dependent 
     $2n\times 2n$ complex, symmetric  matrix such that $\Im M_t$ is positive and  non degenerate.\\
     $A,\ B$ being two classical observables, we have:
$$\left\{A\ ,\ B\right\}= \nabla A\ .\ J\nabla B$$
and
\bea
(\partial_{x}\partial_{\eta}-\partial_{y}\partial_{\xi})^2 A(x,\xi)
 \ B(y, \eta)\vert_{X=Y}&= &\partial^2_{x^2}A\  \partial^2_{\xi^2}B + \partial
 ^2_{x^2}B\ 
 \partial^2_{\xi^2}A- 2 \partial^2_{x \xi}A\  \partial^2_{x \xi} B\nonumber\\
&=&-{\rm Tr}(JA''JB'')
\eea where $A''$ is the Hessian of $A$ (and similarly for $B$). Applying this with 
 $$A(X)= \frac{1}{2}S_{t}X.X, \quad B(X)= \exp (iM_{t}X.X)$$
  we get:
 $$H_{t}\# E_{t} (X) = H_{t}(X)\ E_{t}(X) + JS_{t}X \ .\ M_{t}X\ E_{t}(X)
 + \frac{1}{8}{\rm Tr}(JS_{t}J B'')$$
 However,
 $$\nabla B = 2i B(X)\ M_{t}X$$
 $$(B'')_{jk} = 2i\left((M_{t})_{jk}+2i (M_{t}X)_{j}\ (M_{t}X)_{k}\right)\ B(X)$$
 so that:
 $$
 \frac{1}{8}{\rm Tr}(JS_{t}JB'')= \left(\frac{i}{4}{\rm Tr} (JS_{t}JM_{t}) -\frac{1}{2}
 M_{t}X\ .\ JS_{t}JM_{t}X\right)B(X)
 $$
 Therefore the Ansatz  (\ref{ans})  leads to the equation:
\bea\label{eq2}
 i \dot \alpha(t)-\alpha(t)\dot M_{t}X \ .\ X =\frac{1}{2}\left(S_{t}X\ .\ X
 +M_{t}JS_{t}X\ .\ X -S_{t}JM_{t}X\ .\ X\right)\alpha(t) \nonumber\\
  + \frac{i}{4}\alpha(t){\rm Tr}(\mathcal
 M_{t}\mathcal S_{t})- \frac{1}{2}\alpha(t)M_{t}X\ .\ \mathcal S_{t}
 \mathcal M_{t}X
 \eea
 where we have introduced the Hamiltonian matrices
 $$\mathcal M_{t}:= JM_{t}, \quad \mathcal S_{t}:= JS_{t}$$


Then equation (\ref{eq2}) is equivalent to 
\bea\label{ricc}
\dot{\cal M}_t &=& \frac{1}{2}({\cal M}_t+1){\cal S}_t({\cal M}_t-1)\\
\dot{\alpha}_t &=& \frac{1}{4}{\rm Tr}\left({\cal M}_t{\cal S}_t\right)\alpha_t
\eea
  The first equation is a  Riccati  equation and can be solved with a Cayley transform:
  $$
  {\cal M}_t = (\1-{\cal N}_t)(\1+{\cal N}_t)^{-1}
  $$
  which gives the linear  equation
  $$
  \dot{\cal N}_t = {\cal S}_t{\cal N}_t
  $$
  so we have, recalling that $ {\cal S}_t =  \dot{F}_tF_t^{-1}$, 
  $$
  {\cal N}_t = F_t{\cal N}_0
  $$
  Coming back to ${\cal M}$, we get
  \beq\label{solricc}
  {\cal M}_t = \left(\1+{\cal M}_0-F_t(\1-{\cal M}_0)\right)
  \left(\1+{\cal M}_0+F_t(\1-{\cal M}_0)\right)^{-1}
  \eeq
  Let us now compute the phase term. We introduce
  \beq
  \chi_t^\pm = \1+{\cal M}_0 \pm F_t(\1-{\cal M}_0)
  \eeq
  Using  the following properties
  \bea
   \chi_t^-  &=& \chi_t^+ -2F_t(\1-{\cal M}_0) \\
   {\rm Tr}{\cal S}_t &=& 0\\
   {\cal S}_t &=& \dot{F}_tF_t^{-1}
   \eea
   we have
   \beq
   {\rm Tr}({\cal M}_t{\cal S}_t) = -2{\rm Tr}\left(\dot{\chi}_t^+(\chi_t^+)^{-1})\right)
   \eeq
   so we get
   \beq\label{phase1}
   \alpha_t = \alpha_0\exp\left(-\frac{1}{2}[\log{\rm det}\chi_\bullet^+]_0^t\right),
    \;\; \alpha_0 = 1.
   \eeq
   In formula (\ref{phase1}) the $\log$ is defined by continuity, because we shall
    see that $\chi^+_t$ is always non singular.\\

  Until now we just compute at  the formal level. To make the argument rigorous we state
   some lemmas.\\
  It is convenient here to introduce the following notations.\\
  \noi
  ${\rm sp}_+(2n, \mathbb C)$ is the set of complex, $2n\times 2n$  matrices  ${\cal M}$ such that
   ${\cal M} = JM$ where $M$ is symmetric (that means that ${\cal M}$ is
    a complex  Hamiltonian matrix)
    and  such that $\Im M$ is positive non degenerate.\\
    
    ${\rm Sp}_+(2n, \mathbb C)$ is the set of complex, symplectic, 
  $2n\times 2n$  matrices  ${\cal N}$ such that   the quadratic form
  $$
  z\mapsto \Im\left(\1-\overline{\cal N}^{-1}{\cal N}\right)z\cdot J\overline{z}
  $$
  is positive and non degenerate on $\mathbb C^{2n}$.
  
  \begin{lemma}\label{invert}
  If $F$ is a real  symplectic matrix and ${\cal M} \in   {\rm sp}_+(2n, \mathbb C)$, then 
  $\1+{\cal M} + F(\1-{\cal M})$ is invertible.
 \end{lemma}
 
    \noi {\bf Proof}\\
     Write ${\cal M} = JM$. 
     It is enough to prove that the adjoint  $\1+F^T +{\cal M}^*(\1-F^T)$ is injective.
      But ${\cal M}^* = -\overline{M}J$. So if $z\in \mathbb C^{2n}$ is such that
     $(\1+F^T +{\cal M}^*(\1-F^T))z = 0$ then we get
     \beq\label{inj}
     ((\1+F^T)z-\overline{M}J(\1-F^T)z\cdot J(\1-F^T)\overline{z}) = 0
     \eeq
     But, using that $F$ is symplectic, we have that
     $(1+F)J(1-F^T)=FJ-JF^T$  is symmetric so taking the imaginary part in (\ref{inj}),
     we have 
         $$
     \Im\left(\overline{M}J(\1-F^T)z\cdot J(\1-F^T)\overline{z}\right)= 0.
     $$
     Then using that $\Im M$ is non degenerate, we get successively
     $(\1-F^T)z = 0$, $(\1+F^T)z = 0$ and  $z=0$. \sq

 \begin{lemma}
  Assume that -1 is not an eigenvalue of ${\cal N}$. Then 
 ${\cal N} \in {\rm Sp}_+(2n, \mathbb C)$ if and only if  ${\cal M} \in {\rm sp}_+(2n, \mathbb C)$
  where ${\cal N}$ and ${\cal M}$ are linked by the formula
  $$
  {\cal M} =  (\1-{\cal N})(\1+{\cal N})^{-1}.
  $$
  \end{lemma}

\noi {\bf Proof}\\
Assuming that ${\cal N}+1$ is invertible, and
   $$
  {\cal M} =  (\1-{\cal N})(\1+{\cal N})^{-1}.
  $$
  then we can  easily  see that  ${\cal N}$ is symplectic  if and only if ${\cal M}$ is 
  Hamiltonian.\\
  Now, using  $\overline{\cal N} = {\cal N}^{*,T}$, we get
  $$
  \1 -\overline{\cal N}^{-1}{\cal N} = 
 2(1-\overline{\cal M})^{-1}({\cal M} - \overline{\cal M})(1+{\cal M})^{-1}
 $$
 If $z = (\1+{\cal M})z^\prime$ we have,  
 using $(J(\1 +\overline{\cal M}))^T = -J(\1-\overline{\cal M})$, 
 $$
 (\1 -\overline{\cal N}^{-1}{\cal N})z\cdot J\overline{z}
  = 2(M - \overline M)z^\prime\cdot \overline{z^\prime}.
  $$
  So the  conclusion of  the lemma follows easily from the last equality
  \sq\\
  
 The last lemma has the following  useful   consequence.\\
  Let us start with some ${\cal M}_0\in {\rm Sp} _+(2n, \mathbb C)$
   without the eigenvalue -1. It not difficult to see that
    ${\cal M}_t z = -z$ if and only if ${\cal M}_0u = -u$,
     where  $ u = (\chi_t^+)^{-1}z$. In particular for every time $t$,
      -1 is not an eigenvalue for ${\cal M}_t$.\\
      Furthermore, using that ${\cal N}_t = F_t{\cal N}_0$, we have 
      $\overline{\cal N}_t^{-1}{\cal N}_t = \overline{\cal N}_0^{-1}{\cal N}_0$.
       So we get that the matrix $ {\cal M}_t\in{\rm sp}_+(2n,\mathbb C)$ at every 
        time $t$.\\
      If  ${\cal M}_0$ has the eigenvalue $-1$ it is no more possible to use
   the  Cayley transform but we see that ${\cal M}_t$ is still defined by
     equation (\ref{solricc})(from lemma (\ref{invert}) ) and solves the Riccati equation
      (\ref{ricc}).
     
     Now we want to discuss in more details the phase factor included in the term
      $\alpha_t$ and to consider the limiting case ${\cal M}_0 = 0$
       to compute the Weyl symbol of the propagator $\widehat{U_t^0}$.
       So doing we shall recover the Mehlig-Wilkinson formula,
        including the phase correction Maslov-Conley-Zehnder index).\\
        Let us denote
        $$
        \delta(F_t, {\cal M}_0) =
        {\rm det}\left(\frac{\1+{\cal M}_0 + F_t(\1-{\cal M}_0)}{2}\right).
        $$
        Hence  we have
        $$
        \alpha_t = \exp\left(-\frac{1}{2}\int_0^t
        \frac{\dot\delta(F_t, {\cal M}_0)}{\delta(F_t, {\cal M}_0) }\right)ds.
        $$
        
        \begin{lemma}\label{index}
         Let us consider $t\mapsto F_t $ a path in ${\rm Sp}(n,\mathbb R)$.
         Then  for every ${\cal M}_0\in {\rm sp}_+(n,\mathbb C)$ we have, 
        for the real part:
        $$
       \Re\left[\int_0^t
        \frac{\dot\delta(F_s, {\cal M}_0)}{\delta(F_s, {\cal M}_0) }ds\right]
        =  \log\left(\frac{\vert\delta(F_t, {\cal M}_0)\vert}
        {\vert\delta(F_0, {\cal M}_0)\vert}\right),
        $$
       If $F_t$ is $\tau$-periodic, then
        $$
        \int_0^\tau
        \frac{\dot\delta(F_t, {\cal M}_0)}{\delta(F_t, {\cal M}_0) }ds
         =  2i\pi \nu
        $$
        with $\nu\in \mathbb Z$. Furthermore, $\nu$ is independent on ${\cal M}_0
         \in {\rm sp}_+(2n, \mathbb C)$ and depends only on the homotopy class of
         the closed path $t\mapsto F_t$ in ${\rm Sp}(2n)$.
        \end{lemma}
        
       \noi {\bf Proof}\\
        For simplicity, let us denote
      $\delta(s) = \delta(F_s, {\cal M}_0)$ and
      $$
      h(t) = \int_0^t\dot{\delta}(s)\delta(s)^{-1}ds,\;\; g(t) = {\rm e}^{-h(t)}\delta(t)
      $$
       $g$ is clearly constant in $time$ and $g(0)= \delta(0) = 1$.
        Then we get $\Re(h(t)) = \log\vert\delta(t)\vert$.\\
     In  the periodic case  we have ${\rm e}^{h(\tau)}=1$ so we have
     $$
     \frac{1}{2\pi}\int_0^\tau\frac{\dot\delta(F_t, {\cal M}_0)}{\delta(F_t, {\cal M}_0) }ds
         =  \nu,\;\; \nu\in\mathbb Z.
         $$
         By a simple continuity argument, we see that $\nu$ is invariant by continuous deformation 
          on ${\cal M}_0$ and $F_t$. \sq\\

         We can now compute the Weyl symbol of $\widehat R(F)$
       when ${\rm det}(\1+F) \neq 0$. Let us consider first the case 
       ${\rm det} (\1+F) > 0$. The case ${\rm det}(\1+F) < 0$
       is a little bit more complicated because the identity $\1$ is not in this component.\\
       We start with an arbitrary $C^1$, path $t\mapsto F_t$ going from $\1\ (t=0)$ to 
        $F\ (t=1)$.
      It is known that 
     ${\rm Sp}^+(2n) = \{F\in {\rm Sp}(2n),\; {\rm such\;that}\; {\rm det} (\1+F) > 0\}
     $ is an open  connected subset of ${\rm Sp}(2n)$. 
              So, we can choose a piecewise $C^1$ path $F^\prime_t$
                in  ${\rm Sp}^+(2n)$ going from $\1$ to $F$ and
                 ${\cal M}_0 = i\varepsilon J$. We have, using Lemma(\ref{index}),
                 \beq\label{index2}
                 \Im\left(\int_0^1
        \frac{\dot\delta(F_t, i\varepsilon J)}{\delta(F_t, i\varepsilon J) }dt\right)
 =  2\pi \nu + \Im\left(\int_0^1\frac{\dot\delta(F^\prime_t, i\varepsilon J)}
 {\delta(F^\prime_t, i\varepsilon J )}dt\right)
                 \eeq
But ${\rm det}(\1+F_t^\prime)$ is never 0 on $[0,1]$ and is real;  
                             so if $\varepsilon>0$ is going to zero, the last term in r.h.s  goes to 0
                              and  we get
                 $$
                 \lim_{\varepsilon \searrow 0}
                 \Im\left(\int_0^1
      \frac{\dot\delta(F_t, i\varepsilon J)}{\delta(F_t, i\varepsilon J )}dt\right)
                  =  2\pi \nu
                  $$
                  So we have proved  for the Weyl symbol $R(F, X)$ of $\widehat R(F)$ the 
                   following  Melhig-Wilkinson formula
                 \beq\label{MelWil}
                  R(F, X) = {\rm e}^{i\pi \nu}\vert{\rm det}(\1+F)\vert^{-1/2}
                  \exp\left(-iJ(\1-F)(\1+F)^{-1}X\cdot X\right)
                  \eeq
                  
                  Let us now consider  $F\in {\rm Sp}_-(2n)$
                   where 
                   ${\rm Sp}_-(2n) = \{F\in {\rm Sp}(2n),\; {\rm such\;that}\;
                    {\rm det} (\1+F) < 0\}$.
                   Here we shall replace the identity matrix
                    by 
                    $$
                    F^0_2 = \left(\begin{array}{cc}
             -2&0\\
                      0&-\frac{1}{2}
                      \end{array}\right)
                   $$
                  for $n=1$ and $F^0 = F^0_2\otimes \1_{2n-2}$ for $n\geq 2$
                   where $ \1_{2n-2}$ is the identity in $\mathbb R^{2n-2}$.\\
                    Let us consider a path  connecting   $\1$ to $F^0$ then $F^0$ to
                    $F_1=F$.  Because  ${\rm Sp}(2n)$ is open and connected we can find a path
         in ${\rm Sp}(2n)$ going from  $F^0$ to $F_1=F$ and that part does not contribute
          to the phase by the same argument as above.\\
          Let us consider the model case $n=1$.  The following formula gives an explicit path in 
          ${\rm Sp}(2)$.
                      $$
                    F^\prime_t = \left(\begin{array}{cc}
            \cos t\pi&-\sin t\pi\\
                      \sin t\pi&\cos t\pi
                      \end{array}\right)
                     \left(\begin{array}{cc}
             \eta(t)&0\\
                      0&\frac{1}{\eta(t)}
                      \end{array}\right),
                   $$
where $\eta(t)$ is analytic on a complex neighborhood of $[0, 1]$, $\eta(0)=1$, $\eta(1)=2$,
 $1\leq\eta(t)\leq 2$ for $t\in [0, 1]$.  A simple example is $\eta(t) = 1+t$.
 Then we can compute:
 
 \begin{lemma}
\beq\label{half}
       \lim_{\varepsilon \rightarrow 0}
       \Im\left(\int_0^1
      \frac{\dot\delta(F_t, i\varepsilon J)}{\delta(F_t, i\varepsilon J) }dt\right)
         =  \pi 
      \eeq
 \end{lemma}
 
\noi {\bf Proof}\\
 Let us denote $F_t^\prime=R(t)B(t)$, where $R(t)$ is the rotation matrix
  of angle $t\pi$ and for $\varepsilon\in [0, 1[$, 
  $$
  f_\varepsilon(t) = {\rm det}[(\1-i\varepsilon J)+F_t^\prime(\1-i\varepsilon J)]
  $$
 We have
 $$
 f_0(t)={\rm det}(R(-t) +B(t)) = 2+\cos(t\pi)\left(1+t+\frac{1}{1+t}\right)
 $$                   
$f_0$ has exactly one simple zero $t_1$ on $[0,1],\ f_0(t_1)=0$, $\dot f_0(t_1)\neq 0$).\\
This is easy to see by solving  the equation $\cos t\pi = h(t)$, for   a suitable $h$, with a geometric argument .\\
Then by a standard complex analysis argument (contour deformation) we get 
 the equality (\ref{half}). \sq  \\ 
                   
       So in this case  we have the formula (\ref{MelWil}) for the contravariant Weyl symbol
        of $\hat R(F)$, with index $\nu\in \mathbb Z+1/2$. Summing up the discussion of
         this section we have proved:
    \begin{theorem}\label{contra} We can realize the metaplectic representation $F\mapsto \hat R(F)$
     of the symplectic group ${\rm Sp}(2n,\mathbb R)$ into the unitary group of $L^2(\mathbb R^n)$ by taking for every $F$ a $C^1$- path $\gamma$ going from  $\1(t=0)$ to $F(t=1)$
      and solving explicitly the corresponding quadratic Schr\"odinger equation 
       for the Hamiltonian $H_t(z)=-1/2J\dot F_tF_t^{-1}z\cdot z$. So let us  define $\widehat{R_\gamma}(F)$, 
        the propagator at time $1$ obtained this way.\\
        If $\gamma^\prime$ is another path going from $\1\ (t=0)$ to $F\ (t=1)$
         then there exists an index $N(\gamma,\gamma^\prime)\in \mathbb Z$ such that
         $$
         \widehat{R_\gamma}(F) = {\rm e}^{i\pi N(\gamma,\gamma^\prime)}
         \widehat{R_{\gamma^\prime}}(F)
         $$
         The metaplectic operator $\hat R(F)$ is the two valued unitary operator  
        $\pm\widehat{R_\gamma}(F) $.\\
    Moreover if ${\rm det}(\1+F)\neq 0$,  $\hat R(F) $ has a smooth  contravariant 
     Weyl symbol $R(F, X)$, given by formula (\ref{MelWil}), where $\nu\in\mathbb Z$
      if ${\rm det}(\1+F) >0$ and   $\nu\in\mathbb Z+1/2$
      if ${\rm det}(\1+F) <0$.
      \end{theorem}
     It will be useful  to translate the above theorem for the covariant Weyl symbol.
    Before that we start  to discuss the general case, including  ${\rm det}[F\pm\1]=0$. 
     The Weyl symbol  of $\hat R(F)$ may be singular, 
        so it is easier to analyse it using coherent states (for our application it is exactly what we need).\\
          Let us recall that
        \beq\label{quadcov}
           R^\#(F, X) = (2\pi\hbar)^{-n}\int_{\mathbb R^{2n}}\langle\varphi_{z+X}\vert\hat R(F)\varphi_z\rangle {\rm e}^{-\frac{i}{2}\sigma(X,z)}dz
           \eeq
         Let us denote $\hat U_1 = \hat R(F)$ and $U_1^\varepsilon$ the contravariant Weyl symbol at time 1
          constructed above, such that $ U_0^\varepsilon(X) = {\rm e}^{-\varepsilon\vert X\vert^2}$.\\
          For every $\varepsilon > 0$ we have computed the following  formula for the contravariant Weyl symbol:
          \bea  
     U^\varepsilon(X) &=& \alpha^\varepsilon\exp\left(iM^\varepsilon X\cdot
      X\right), \;{\rm where}\nonumber\\
     M^\varepsilon(X) &=& = -J(\1+i\varepsilon J-F(\1-i\varepsilon J))(\1+i\varepsilon J+F(1-i\varepsilon J))^{-1}\\
     \alpha^\varepsilon &=& {\rm det}(\1+i\varepsilon J+F(\1-i\varepsilon J))^{-1/2}          
          \eea
            At the end  we get the result by taking the limit:
           \beq
           \lim_{\varepsilon\rightarrow +\infty}\langle\varphi_{z+X}\vert\hat U_1^\varepsilon\varphi_z\rangle
            = \langle\varphi_{z+X}\vert\hat R(F)\varphi_z\rangle
            \eeq
        The computation uses the following formula for the Wigner function, $W_{z, z+X}(Y)$,  of the pair
        $(\varphi_z, \varphi_{z+X})$.
        \beq
        W_{z, z+X}(Y) = 2^{2n}\exp\left(-\left\vert Y- z-\frac{X}{2}\right\vert^2
        -i\sigma(X,Y-\frac{z}{2}  )\right)
        \eeq
        So, we have to compute the following Fourier-Gauss integral
        \beq
        \langle\varphi_{z+X}\vert\hat U_1^\varepsilon\varphi_z\rangle
= 2^{2n}(2\pi)^{-n}\alpha^\varepsilon\int_{\mathbb R^{2n}}dY
\exp\left(iM^\varepsilon Y\cdot Y-\left\vert Y- z-\frac{X}{2}\right\vert^2-
i\sigma(X,Y-\frac{z}{2})\right)
\eeq
    So we get
    \beq
            \langle\varphi_{z+X}\vert\hat U_1^\varepsilon\varphi_z\rangle
= 2^n\left({\rm det}(\1-iM^\varepsilon)\right)^{-1/2}\alpha^{\varepsilon}
.
\eeq
\beq
\times\exp\left(-\left\vert z+\frac{X}{2}\right\vert^2+\frac{1}{2}i\sigma(X,z)
+(\1-iM^\varepsilon)^{-1}(z+\frac{X-iJX}{2})\cdot (z+\frac{X-iJX}{2})\right)
\nonumber
\eeq
Now we can compute the limit when $\varepsilon\searrow 0$. We have
\beq
\1-iM^\varepsilon = \1+iJ(\1+i\varepsilon J-F(\1-i\varepsilon J))(\1+i\varepsilon J-F(\1-i\varepsilon J))^{-1}
\eeq
Then we get, using that  $(\1+F+iJ(\1-F))$ is invertible (see Lemma \ref{invert}), 
\beq
\lim_{\varepsilon\rightarrow 0}(\1-iM^\varepsilon )^{-1} = (\1+F)(\1+F+iJ(\1-F))^{-1}
\eeq
and 
\beq
\lim_{\varepsilon\rightarrow 0}{\rm det}(\1-iM^\varepsilon )^{-1/2}
\alpha^\varepsilon = 
  \left({\rm det}(\1+F+iJ(\1-F)\right)^{-1/2}
 \eeq
 So, finally, we have proved the following
 \begin{proposition}\label{matelem}
 The matrix elements of $\widehat R(F)$ on coherent states
 $\varphi_z$, are given by
  the following formula:
 \bea\label{gaussmet}
 \langle\varphi_{z+X}\vert\hat R(F)\varphi_z\rangle =
 2^n\left({\rm det}(\1+F+iJ(\1-F)\right)^{-1/2}.\nonumber\\
  \times\exp\left(-\left\vert z+\frac{X}{2}\right\vert^2+\frac{1}{2}
  i\sigma(X,z)+K_F  (z+\frac{X-iJX}{2})\cdot (z+\frac{X-iJX}{2})\right)
  \eea
   where
   \beq
   K_F = (\1+F)(\1+F+iJ(\1-F))^{-1}
  \eeq
  \end{proposition}
  Now we  can compute the distribution covariant symbol of 
  $\hat R(F)$  by plugging formula  (\ref{gaussmet}) in formula (\ref{quadcov}).\\
   Let us begin  with the regular case ${\rm det}(\1-F)\neq 0$.
   \begin{corollary}\label{melwil}
        If ${\rm det}(\1-F)\neq 0$,  the covariant Weyl symbol of  $\widehat{R_\gamma}(F)$
        is computed by the formula:
        \beq\label{MelWilcov}
                  R^\#(F, z) = {\rm e}^{i\pi \mu}\vert{\rm det}(\1-F)\vert^{-1/2}
                  \exp\left(-\frac{i}{4}J(\1+F)(\1-F)^{-1}z\cdot z\right)
                  \eeq
                 where
                 $ \mu = \bar\nu+\frac{n}{2}$, 
                $\bar\nu\in\mathbb Z$ is an index computed below in formulas (\ref{ind1}),
                (\ref{ind2}).
                           \end{corollary}
        
      \noi {\bf Proof}\\
        Using Proposition (\ref{matelem})  and formula (\ref{quadcov}), we have to compute a Gaussian
         integral with a  complex, quadratic, non degenerate covariance matrix (see \cite{ho1}).\\
         This covariance matrix is $K_F-\1$ and we have clearly
         $$
         K_F-\1 = -iJ(\1-F)(\1+F+iJ(\1-F))^{-1} = -(\1-i\Lambda)^{-1}
         $$
         where $\Lambda = (\1+F)(\1-F)^{-1}J$ is a real symmetric matrix. So we have
         \beq
         \Re(K_F-\1) = -(\1+\Lambda^2)^{-1},\;\; \Im(K_F-\1) =  -\Lambda(\1+\Lambda^2)^{-1}
         \eeq
         So that $\1-K_F$ is in  the Siegel space $\Sigma_{2n}$ and Theorem (7.6.1) of \cite{ho1} can be applied.
         The only serious problem is to compute the index $\mu$.   \\
         Let us define a path of  $2n\times 2n$ symplectic matrices  as follows:
          $G_t = {\rm e}^{t\pi J_{2n}}$ if  ${\rm det}(\1-F) > 0$ and:\\
           $G_t =  G^2_t\otimes {\rm e}^{t\pi J_{2n-2}}$ if ${\rm det}(\1-F) < 0$, where
         $$
                    G^2 = \left(\begin{array}{cc}
             \eta(t)&0\\
                      0&\frac{1}{\eta(t)}
                      \end{array}\right)
                   $$
                where $\eta$ is a  smooth function on $[0, 1]$ such that
                 $\eta(0)=1$, $\eta(t) > 1$ on $]0, 1]$ and   where $J_{2n}$ is the  $2n\times2n$ 
                  matrix defining the symplectic matrix on the Euclidean space
              $\mathbb R^{2n}$.\\
             $G_1$ and $F$ are in the same connected component of 
             ${\rm Sp}_\star(2n)$ 
              where ${\rm Sp}_\star(2n) = \{F\in {\rm Sp(2n)},\; {\rm det}(\1-F)\neq0\}$. So we can consider a path
               $s\mapsto F^\prime_s$ in ${\rm Sp_\star(2n)}$ such that $F^\prime_0=G_1$ and $F^\prime_1=F$.\\
              Let us consider the following ``argument  of determinant" functions  for families of complex matrices.
            \bea
              \theta[F_t] &=& \arg_c[{\rm det}(\1+F_t+iJ(\1-F_t)]\\
              \beta[F] &=& \arg_+[{\rm det}(\1-K_F)^{-1}]
              \eea
              where $\arg_c$ means that $t\mapsto  \theta[F_t]$ is continuous in $t$ and $\theta[\1]=0$ ($F_0=\1$),
               and $ S\mapsto  \arg_+[{\rm det}(S)]$ is  the analytic determination  defined on the Siegel space
                $\Sigma_{2n}$ such that $ \arg_+[{\rm det}(S)] =0$ if $S$ is real (see \cite{ho1}, vol.1, section (3.4)).\\
                With these  notations we have 
                \beq\label{ind1}
                \mu = \frac{\beta[F]-\theta[F]}{2\pi}.
                \eeq
                Let us consider first the case ${\rm det}(\1-F)> 0$.\\
                Using that $J$ has the spectrum  $\pm i$, we get: ${\rm det}(\1+G_t+iJ(\1-G_t))=4^n{\rm e}^{nt\pi i}$
                 and $\1-K_{G_1} = \1$.  \\
                 Let us remark that ${\rm det}(\1-K_F)^{-1} = {\rm det}(\1-F)^{-1}{\rm det}(\1-F+iJ(\1+F))$. 
                  Let us introduce $\triangle(E,{\cal M}) = {\rm det}(\1-E+{\cal M}(\1+E))$
                   for  $E\in{\rm Sp}(2n)$ and ${\cal M}\in {\rm sp}_+(2n,\mathbb C)$. Let consider 
                    the closed  path ${\cal C}$ in ${\rm Sp}(2n)$ defined by adding $\{G_t\}_{0\leq t\leq 1}$
                     and
                     $\{F^\prime_s\}_{0\leq s\leq 1}$. We denote by $2\pi\bar\nu$ the variation of the argument for
                     $\triangle(\bullet, {\cal M})$ along $ {\cal C}$.  Then we  get easily
                      \beq\label{ind2}
                      \beta(F) = \theta[F] + 2\pi\bar\nu + n\pi,\; n\in\Z.
                      \eeq
                      When  ${\rm det}(\1-F) < 0$, by an explicit computation, we find $\arg_+[{\rm det}(\1-K_{G_1})] = 0$.
                      So we can conclude as above. 
                      \sq

              Assume now  that $F$ has the eigenvalue 1 with some multiplicity $2d$. 
              We want to compute the temperate distribution $R^\sharp(F)$ as a limit
               of $R^\sharp(F^\varepsilon)$ where ${\rm det}(\1-F^\varepsilon)\neq 0$, $\forall \varepsilon>0$.\\
               Let us introduce the generalized eigenspace
                $\di{{\cal E}^\prime = \bigcup_{j\geq 1}\ker(\1-F)^j}$ (dim[${\cal E}^\prime$]=2d).
                and ${\cal E}^{\prime\prime}$ its symplectic orthogonal in $\mathbb R^{2n}$.  
                We denote $F^\prime$ the restriction of $F$ to 
                ${\cal E}^\prime$ and $F^{\prime\prime}$ the restriction to ${\cal E}^{\prime\prime}$. We also denote by
                $J^\prime$ and $J^{\prime\prime}$ the symplectic applications defined by the restrictions of the symplectic
                 form $\sigma$: $\sigma(u,v)= J^\prime u\cdot v$, $\forall u, v\in{\cal E}^\prime$ and the same
                  for $J^{\prime\prime}$. Let us  introduce the Hamiltonian maps
                  $ L^\prime = (\1-F')(\1+F')^{-1}$ and $L^{\prime,\varepsilon} = L^\prime-\varepsilon J^\prime$.
                  
                   It is clear that ${\rm det}(L^{\prime,\varepsilon} - \1) \neq 0$ for 
       $0< \varepsilon$ small enough, so we can define 
       $ F^{\prime,\varepsilon} = 
       (\1 +L^{\prime,\varepsilon})(\1 - L^{\prime,\varepsilon})^{-1}$.
        Let us remark that 
           \beq\label{symmat}
        Q^{\prime,\varepsilon} :=  
     J^\prime(F^{\prime,\varepsilon} -\1)^{-1}(F^{\prime,\varepsilon} + \1) = 
     -(L^\prime J^\prime + \varepsilon)^{-1}
     \eeq
 is a  symmetric non degenerate matrix in ${\cal E}^\prime$ defined for  every $\varepsilon > 0$.

      \begin{lemma}
       We have the following properties.\\
       1) $F^{\prime,\varepsilon}$ is symplectic.\\
       2)
$\di{\lim_{\varepsilon\rightarrow 0}F^{\prime,\varepsilon}} =  F^\prime.$\\
       3) For
$\varepsilon \neq 0$, small enough, ${\rm det}(F^{\prime, \varepsilon} - \1)
\neq 0$.
      \end{lemma}
\noi {\bf Proof}. \\ 1) comes from the fact that
$L^{\prime,\varepsilon}$ is Hamiltonian.\\  2) is clear. \\
 For 3), let us
assume that $F^{\prime,\varepsilon} u = u$. Then we have 
  $L^{\prime,\varepsilon}u
= 0$ hence $J^\prime L^\prime u = \varepsilon u$. Now, 
  choose $0 <\varepsilon < {\rm dist}\{0, 
 {\rm  spec}(J^\prime L^\prime)\backslash\{0\} \}$
then we have $u = 0$. \\
Finally we define $F^\varepsilon =
F^{\prime,\varepsilon} \otimes F^{\prime\prime,\varepsilon}$. 
It is clear that $F^\varepsilon$
satisfies also properties 1), 2), 3) of the 
above lemma with $F^\varepsilon $ in place
of $F^{\prime,\varepsilon}$. 
\sq\\
           
   \begin{proposition}
Under the above assumption, the covariant symbol of $\hat R(F)$ has the following form:
\bea
R^\#(F,z_1,z_2,z^{\prime\prime})= e^{i\pi\mu_1}\vert{\rm det}(\1-F^{\prime\prime})\vert ^{-1/2}\delta(z_{1})\nonumber\\
\times\exp\left(\frac{i}{4}J(\1+F)(F-\1)^{-1}(z_2+z^{\prime\prime})\cdot (z_2+z^{\prime\prime})\right)
\eea
where $z:=((z_{1},z_2), z^{\prime\prime})$ is the decomposition of the phase-space for which
$F = F'\otimes F^{\prime\prime}$ and $z^\prime=z_1+z_2$, with $z_2\in{\rm Im}(F'-\1)$, $z_1\in{\rm Im}(F'-\1)^\perp$,
 the orthogonal complement in ${\cal E}^\prime$ for the Euclidean scalar product.
   $\delta(z_{1})$ denotes the Dirac mass at point $z_{1}=0$. $\mu_1\in \mathbb Z +1/2$ is  given
   as follows: $\mu_1 = \mu^{\prime\prime}+\frac{ {\rm sg}^+Q^\prime}{4}$ where  $\mu^{\prime\prime}$ is the limit of the $\mu$ index 
    for $F^\varepsilon$, computed in (\ref{MelWilcov}), 
     and ${\rm sg}^+Q^\prime$ is the limit for $\varepsilon \rightarrow 0$ of
     the signature of $Q^{\prime,\varepsilon}$, defined in (\ref{symmat}).
     \end{proposition} 

\noi{\bf Proof}\\ 
We use the same kind of computation as for Corollary (\ref{melwil}). The new factor comes from the contribution
 of ${\cal E}'$. So we can forget  ${\cal E}^{\prime\prime}$. So we assume that ${\cal E}'=\mathbb R^{2n}$ and
  and we forget the superscripts '.  We have to compute
  the limit in the distribution sense of $R^\sharp(F^\varepsilon)$ which was computed in Corollary (\ref{melwil}).\\
  Let us consider a test function $f\in{\cal S}(\mathbb R^{2n})$ and its Fourier transform $\tilde f$.
  Using Plancherel  formula, we have
 \bea\label{Four}
    \int_{\mathbb R^{2n}}\exp\left(-\frac{i}{4}J(\1+F^\varepsilon)(\1-F^\varepsilon)^{-1}z\cdot z\right) f(z)dz = \nonumber\\
    2^n(\pi)^{2n}\vert{\rm det}(Q^\varepsilon)\vert^{-1/2}{\rm e}^{{\rm sg}(Q^\varepsilon)\frac{i\pi}{4}}
    \int_{\mathbb R^{2n}} \exp\left(-i(\1+F^\varepsilon)^{-1}(\1-F^\varepsilon)J\zeta\cdot \zeta\right)\tilde f(\zeta)d\zeta
  \eea
So we get the result  by taking the limit for $\varepsilon\rightarrow 0$ in (\ref{Four}).\\
Let us remark that we have used that $J^\prime(\1+F)(\1-F)^{-1}$ is an isomorphism
from $\ker(\1-F^\prime)^\perp$ onto itself.
\sq
          
  In our paper \cite{coro2}  the leading term for the return probability and for fidelity on coherent states
   is computed  with a Gaussian exponential defined by the quadratic form
    which was defined in Proposition \ref{matelem}. 
  \beq
  \gamma_F(X) :=  \frac{1}{4}\left(K_F(\1-iJ)X\cdot(\1-iJ)X-\vert X\vert^2\right)
   \eeq
 In our application we shall have $X = z_t-z$ where $t\mapsto z_t$ is the classical path
  starting from $z$. So we need  to estimate the argument in the exponent
   of formula (\ref{gaussmet})  for $z =0$.
  \begin{lemma}
  we have, $\forall X\in\mathbb R^{2n}$, 
  \beq
  \Re(\gamma_F(X)) \leq -\frac{\vert X\vert^2}{2(1+s_F)},
  \eeq
  where $s_F$ is the largest eigenvalue of $FF^T$ ($F^T$ is the transposed matrix
   of $F$) .
    \end{lemma}     
  {\bf Proof}\\
  Let us begin by assuming that ${\rm det}(\1+F)\neq 0$. Then we have
  $$
  K_F = (\1+iN)^{-1},\;{\rm where}\;\; N = J(\1-F)(\1+F)^{-1}.
  $$
  So we can compute
  $$
  \Re (K_F) = (\1+N^2)^{-1} = K_FK_F^\star \;\;{\rm and}\;\; \Im(K_F) = -N(\1+N^2)^{-1}.
  $$               
    So, we get, 
    \beq
    \gamma_F(X) = \frac{1}{4}\left((\1+JN)K_FK_F^\star(\1-NJ)X\cdot X-2\vert X\vert^2\right)
    \eeq
    By definition of $K_F$, we have
    \beq
    (\1+JN)K_F = 2\left((\1+iJ)F^{-1} + \1-iJ\right)^{-1}
       \eeq
    Le us denote $T_F = ((\1+iJ)F^{-1} + \1-iJ)^{-1}$. We have, using
     that $F$ is symplectic, 
    $$
    T_F^{\star, -1}T_F^{-1} = 2(F^{-1, T}F^{-1} +\1).
    $$
   hence we get
   $$
   T_FT_F^\star = (2(F^{-1, T}F^{-1} +\1))^{-1}
   $$ and the conclusion of the lemma follows for ${\rm det}(\1-F)\neq 0$
    hence for every symplectic matrix $F$ by continuity.\sq
    
    \vspace{1cm}
    \noi
    {\tt  Acknowledgement} \\ {\em We begun this work by discussions around the Mehlig-Wilkinson formula and application 
     to  the Loschmidt echo at Mittag-Leffler Institute, in fall 2002.
     We thank the institute  and  the organizers of the semester  on  Partial Differential Equations and Spectral Theory, 
 A. Laptev, V. Guillemin and B. Helffer  for their hospitality 
 and exceptionnal  working surroundings.}

\end{document}